\documentclass[conference]{IEEEtran}
\IEEEoverridecommandlockouts

\usepackage{cite}
\usepackage{amsmath,amssymb,amsfonts}
\usepackage{algorithmic}
\usepackage{graphicx}
\usepackage{textcomp}
\usepackage{xcolor}
\usepackage{url}
\usepackage{balance}
\usepackage{booktabs}
\usepackage{array}
\usepackage{colortbl}

\def\BibTeX{{\rm B\kern-.05em{\sc i\kern-.025em b}\kern-.08em
    T\kern-.1667em\lower.7ex\hbox{E}\kern-.125emX}}

\title{Auditing Blockchain Innovations: Technical Challenges Beyond Traditional Finance}

\author{
\IEEEauthorblockN{Shayan Eskandari\IEEEauthorrefmark{1}}
\IEEEauthorblockN{Leid Zejnilovic\IEEEauthorrefmark{1}}
\IEEEauthorblockN{Jeremy Clark\IEEEauthorrefmark{2}}
\IEEEauthorblockA{\IEEEauthorrefmark{1}Nova School of Business and Economics, Universidade NOVA de Lisboa,\\
Campus de Carcavelos, 2775-405 Carcavelos, Portugal\\
\{shayan.eskandari, leid.zejnilovic\}@novasbe.pt}
\IEEEauthorblockA{\IEEEauthorrefmark{2}Concordia University, Montreal, Canada\\
j.clark@concordia.ca}
}

\begin{document}

\maketitle

\begin{abstract}
Blockchain technology introduces asset types and custody mechanisms that fundamentally break traditional financial auditing paradigms. This paper presents an autoethnographic analysis of cryptoasset auditing challenges, build on top of prior research on a comprehensive framework addressing  existence, ownership, valuation, and internal control verification. Drawing from lived experience implementing blockchain systems as an engineer, smart contract auditor, and CTO of a publicly traded cryptoasset firm, we demonstrate how autoethnographic methodology becomes necessary for understanding technical complexities that external analysis cannot capture. Through detailed examination of token airdrops, multi-signature smart contracts, and real-time on-chain reporting, we provide experimental approaches and common scenarios that auditing firms can analyze to address blockchain innovations currently considered technically insurmountable.
\end{abstract}


\section{Introduction}

The \$4 trillion blockchain industry faces a critical barrier: major auditing firms refuse to certify companies holding significant cryptoassets, leading to cease trade orders and blocked access to capital markets~\cite{pimentel2021systemizing,posadzki2019crypto}. This reluctance stems from blockchain innovations that introduce fundamental paradigm shifts requiring entirely new audit procedures.

Unlike incremental technological changes accommodated within traditional procedures, blockchain creates new asset categories: tokens appearing without purchase transactions (airdrops), ownership structures without private key control (multi-signature smart contracts), and continuous financial reporting from immutable ledgers~\cite{dai2017toward,rozario2019reengineering}. These innovations break traditional audit frameworks designed for centralized systems with clear ownership documentation~\cite{hsieh2022issues} and authorized (financial) statements.

Recent industry examples demonstrate the severity of this challenge. Several crypto companies, including those operating in regulated environments, were placed on cease trade orders when auditors abruptly stepped down, leaving firms unable to find replacement auditors~\cite{boulianne2020risks,globemail2019crypto}. Even major exchanges like Binance have struggled to obtain audit certifications for proof-of-reserves procedures~\cite{wagner2022crypto}.

\textbf{The Necessity of Autoethnographic Methodology.} Traditional external audit analysis fails to capture the subtle technical dependencies that impact financial verification procedures. Consider multi-signature smart contracts: external observation might classify these as "complex custody arrangements," but only hands-on implementation reveals that ownership verification requires analyzing contract source code (often bytecode), validating governance mechanisms, and assessing upgrade risks—procedures with no traditional audit equivalent.

This paper employs autoethnographic methodology~\cite{ellis2011autoethnography} to analyze blockchain audit challenges through lived professional experience across multiple industry roles. This approach is justified by the technical impossibility of understanding these systems without implementation experience~\cite{kothari2008research,yin2013case}.

\section{Methods}

We adopt an applied research approach with theory-building perspective, addressing practical issues as they emerge and suggesting potential solutions~\cite{toffel2016enhancing}. This methodology combines systematic analysis with personal professional experience to bridge the knowledge gap between technical blockchain expertise and financial auditing requirements.

\textbf{Professional Context:} The analysis draws from lived experience of one of the authors and their progression through multiple blockchain ecosystem roles: (1) \textit{Blockchain Engineer} providing foundational understanding of cryptocurrency transaction flows and key management; (2) \textit{Smart Contract Security Auditor} revealing technical complexities requiring domain-specific expertise; (3) \textit{Chief Technology Officer} exposing the disconnect between technical implementation and financial reporting requirements through direct auditor interaction.

\textbf{Case Study Methodology:} We employ paradigmatic case research~\cite{cooper2008case} to examine specific blockchain innovations that fundamentally challenge traditional audit frameworks. This approach provides information about situations auditors face when attempting to verify emerging cryptoasset types and custody mechanisms. Three illustrative cases are analyzed: (1) token airdrop existence verification challenges; (2) self-custody ownership evolution; and (3) real-time blockchain-based financial reporting. Each case study combines technical implementation details with autoethnographic analysis of auditor interactions, revealing systematic gaps between traditional audit procedures and blockchain innovation requirements. The research team combines "complete participation" for technical implementation with "observer participant" analysis, balancing insider insights with external academic perspective~\cite{vinten1994participant}.

\begin{table*}[t]
\centering
\caption{Autoethnographic Data Sources Overview: Professional Experience}
\label{tab:autoethno_data}
\footnotesize
\definecolor{lightgray}{gray}{0.95}
\begin{tabular}{p{3.7cm}p{3.7cm}p{1.6cm}p{7cm}}
\toprule
\textbf{Experience Category} & \textbf{Quantitative Scope} & \textbf{Time Period} & \textbf{Autoethnographic Analysis Focus \& Insights} \\
\midrule
\rowcolor{white}
Blockchain Engineering: Bitcoin~ATM \& Cloud~Wallet Management & 
\begin{minipage}[t]{3.7cm}
• 1000+~ATM~deployments\\
• \$1B+~transaction~volume
\end{minipage} & 
2015--2018 & 
\begin{minipage}[t]{7.0cm}
• Understanding Bitcoin operational complexity\\
• Key management practical challenges\\
• Gaps between theoretical security \& implementation requirements
\end{minipage} \\[1.5ex]

\rowcolor{lightgray}
Smart~Contract~Security Auditing: DeFi~protocol assessments & 
\begin{minipage}[t]{3.7cm}
• 50+~protocol~audits\\
• 4000+~total~audit~hours \\
• 10~conference~presentations
\end{minipage} & 
\begin{minipage}[t]{1.4cm}
2018--2021\\
2023--2025
\end{minipage} & 
\begin{minipage}[t]{7.0cm}
• Technical complexity assessment methodologies\\
• DeFi vulnerability identification\\
• Code analysis procedures \& audit frameworks\\
• Domain-specific expertise requirements
\end{minipage} \\[1.5ex]

\rowcolor{white}
Chief~Technology~Officer: Publicly-traded cryptoasset~firm & 
\begin{minipage}[t]{3.7cm}
• \$100M+~AUM\\
• 2~annual~audit~engagements\\
• 6~quarterly~financial~statements\\
• 15~airdrop~claims \\
• 5~regulatory~workshops
\end{minipage} & 
2021--2023 & 
\begin{minipage}[t]{7.0cm}
• Direct auditor interaction experiences\\
• Financial reporting integration challenges\\
• Institutional custody complexities\\
• Auditor knowledge gaps and cultural resistance \\
• DeFi risk assessment challenges
\end{minipage} \\[1.5ex]

\rowcolor{lightgray}
Multisignature~Implementation: Organizational~custody & 
\begin{minipage}[t]{3.7cm}
• 10+~multisig~deployments\\
• 12~governance~configurations
\end{minipage} & 
2019--2025 & 
\begin{minipage}[t]{7.0cm}
• Governance mechanism design trade-offs\\
• Control framework development insights\\
• Technical custody evolution impact on audits
\end{minipage} \\[1.5ex]

\bottomrule
\end{tabular}
\end{table*}

Table~\ref{tab:autoethno_data} summarizes the autoethnographic data sources informing this analysis. We employ evocative autoethnography~\cite{ellis2011autoethnography} to capture lived experience of navigating technical-financial boundaries that external analysis cannot access, revealing implicit knowledge embedded in professional practice.

\textbf{Autoethnographic Validation:} Following established practice for single-researcher autoethnography, we employed systematic triangulation through the industry engagement documented in Table~\ref{tab:autoethno_data}. Conference presentations (10+ across security auditing contexts), regulatory workshops (5+ with compliance teams), and direct auditor engagements (2 annual audits and 6 quarterly reporting cycles) provided opportunities to test emerging insights against the experiences of other blockchain-audit interface practitioners. This validation process revealed consistent patterns across different organizational contexts, supporting the transferability of our autoethnographic findings~\cite{ellis2011autoethnography}. The framework's resonance with industry practitioners—from technical auditors to regulatory compliance teams—provided external validation of insights derived from personal professional experience spanning over 4000 hours of protocol auditing and \$100M+ in assets under management.

\section{Cryptoasset Audit Framework}

\begin{figure}[t]
    \centering
    \includegraphics[width=\columnwidth]{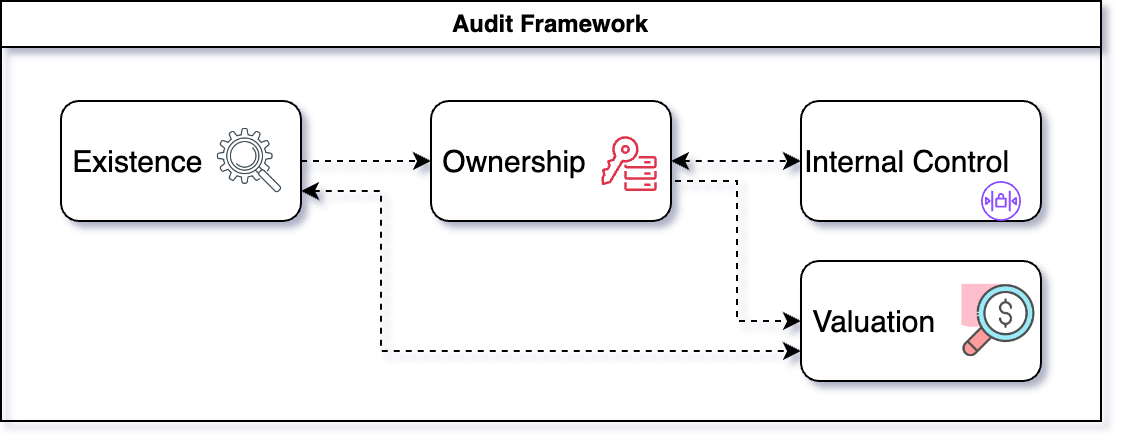}
    \caption{Cryptoasset Audit Framework: A cryptoasset exists when it has material value and is owned by an entity. Ownership requires internal control to keep assets safe and accessible.}
    \label{fig:audit_framework}
\end{figure}

Our autoethnographic analysis organizes blockchain audit challenges into four interconnected categories (Figure~\ref{fig:audit_framework}), extending prior work~\cite{pimentel2021systemizing} with practical implementation insights.

\subsection{Existence Challenges}

\textbf{Technical Innovation Impact:} Blockchain creates assets through mechanisms that have no traditional acquisition equivalents. Token airdrops create assets appearing in company wallets without purchase transactions, delivery mechanisms, or counterparty relationships~\cite{allen2023airdrop}.

\textbf{Reliability Assessment:} Not all blockchain-based systems provide equivalent existence assurance. Auditors must evaluate consensus mechanism robustness, validator diversity, community support, cryptographic security~\cite{dunn2021bitcoin}, and system implementation. Bitcoin and Ethereum enjoy extensive validation, while obscure tokens may reside on unreliable networks.

\textbf{Fork and Airdrop Complications:} Existing accounting standards cannot accommodate non-reciprocal asset transfers~\cite{webb2018fork}. The Ethereum merge or Bitcoin Cash fork, resulted in the creation of separate holdings on the blockchain forks~\cite{themerge,bitcoincashabcFork}, while Optimism airdrops required complex claiming procedures with technical dependencies~\cite{optimismgithub}.

\subsection{Ownership Verification}

\textbf{Self-Custody Challenges:} Absent legal registers or official documents, auditors must rely on cryptographic proof systems. Ownership verification requires temporal considerations—proving control at fiscal year-end rather than audit date—and understanding the distinction between access and ownership~\cite{vincent2020challenges}.

\textbf{Smart Contract Complexity:} Multi-signature contracts implement governance mechanisms, upgrade procedures, and access controls with no traditional custody equivalent~\cite{eyal2022cryptocurrency}. Unlike single key accounts (EOA), smart contracts require code analysis and governance validation~\cite{zhao2018secure}.

\textbf{Custodial Arrangements:} Third-party custody requires evaluating Service Organization Control (SOC) reports, but these often inadequately address cryptoasset-specific controls~\cite{bdosocreports}. Auditors must assess custodian segregation practices and technical implementation details.

\subsection{Valuation Complexities}

\textbf{Market Fragmentation:} Cryptoasset markets operate continuously across jurisdictions with fragmented liquidity and geographical price variations~\cite{kroeger2017law}. IFRS requirements for principal market identification become complex when trading occurs on multiple decentralized platforms~\cite{ifrs13fairvalue}.

\textbf{Fungibility Issues:} Digital assets introduce fungibility challenges unknown in traditional finance. "Dirty" coins with histories involving in hacking incidents or privacy protocols like Tornado Cash may trade at discounts, creating valuation complexities~\cite{ofactornadocash,pernice2019monetary}.

\subsection{Internal Control Innovation}

\textbf{Technical Control Requirements:} Effective cryptoasset internal controls require understanding hardware security modules (HSM), multi-signature threshold selection, secure key generation, and backup procedures.~\cite{c4ccssa}.

\textbf{Control Trade-offs:} More secure controls (multi-signature wallets) complicate ownership verification, while simpler approaches (single keys) create single points of failure. Auditors must evaluate these technical trade-offs without established time-tested frameworks~\cite{gaggioli2019middleman}.

\section{Auditing Challenges on Blockchain Innovations}

\subsection{Case Study 1: Airdrop Existence Verification}

\begin{figure}[t]
    \centering
    \includegraphics[width=\columnwidth]{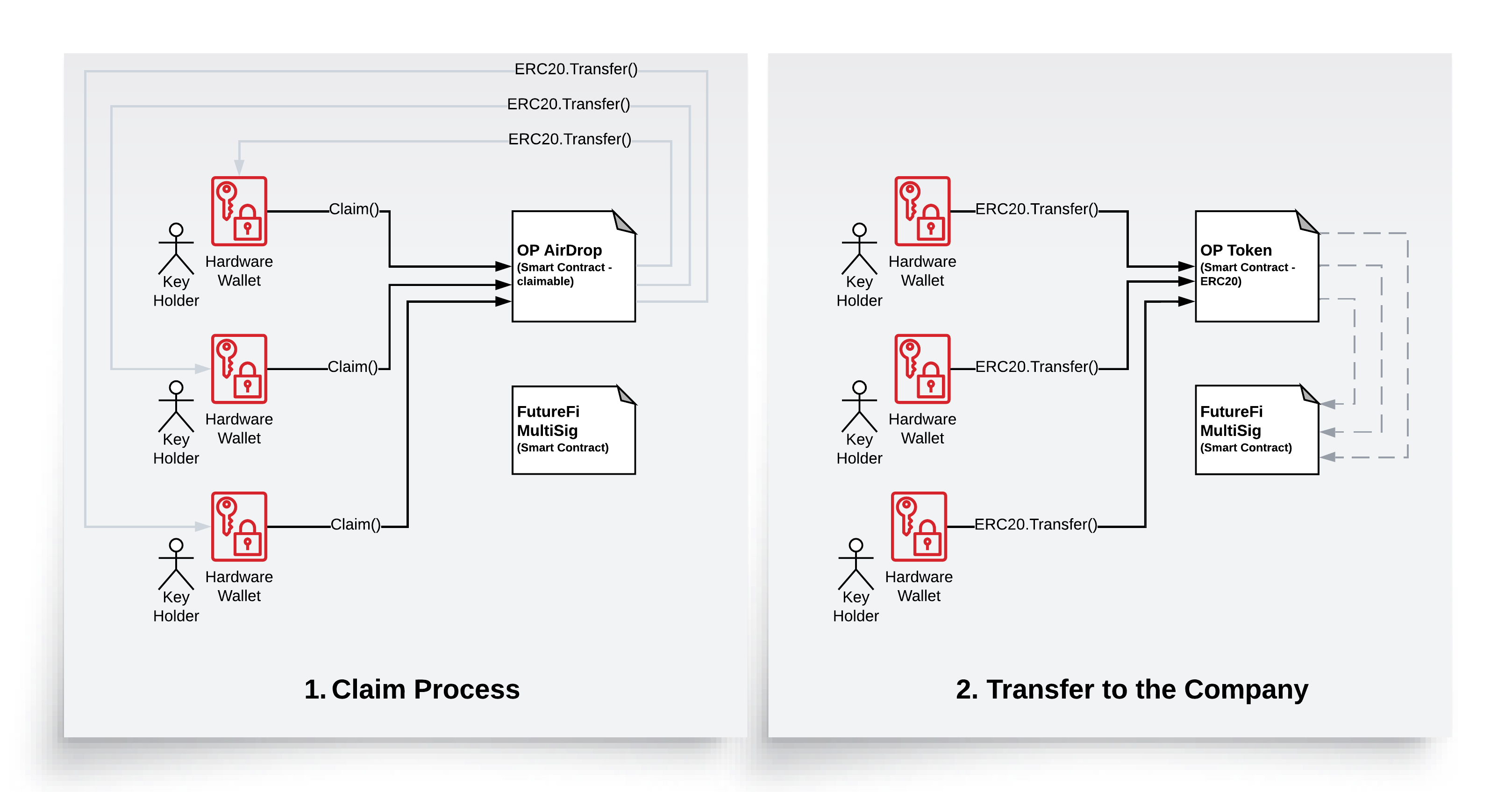}
    \caption{Case Study 1 - Optimism Airdrop First Claim Workflow for Multi-signature Custody}
    \label{fig:opairdrop_workflow}
\end{figure}

\textbf{Technical Challenge:} The Optimism airdrop required individual claiming transactions from each multi-sig keyholder, with gas costs paid separately, creating complex technical dependencies (Figure~\ref{fig:opairdrop_workflow})~\cite{optimismgithub}.

\textbf{Autoethnographic Analysis:} Managing 15 airdrop claims across multiple protocols revealed that these events often occur without notification, require protocol-specific claiming procedures, and involve technical risks that can jeopardize existing holdings. The moment I realized traditional auditing had no framework for airdrops came during our first quarterly audit when auditors requested "purchase documentation" for tokens that had simply appeared in our wallets. This disconnect—between their need for acquisition evidence and the reality of permissionless token distribution—revealed the fundamental inadequacy of traditional audit frameworks. Multi-phase airdrops (Optimism had two phases) require different procedures for each phase, with gas costs varying by network congestion and claiming complexity. The lived experience of repeatedly explaining blockchain mechanics to confused auditors across 6 quarterly reporting cycles illustrated the persistent cultural and knowledge gap that autoethnographic analysis can bridge~\cite{ellis2011autoethnography}. Each airdrop presented unique verification challenges: some required snapshot proofs, others involved staking requirements, and several had time-sensitive claiming windows that created additional audit timing complications.

\textbf{Auditing Challenges:} Traditional audit procedures cannot verify these technical dependencies. Instead, auditors must develop experimental approaches to confirm airdrop existence, ownership, and claiming procedures, as well as conduct gas cost-based analysis for valuing the airdropped tokens.

\subsection{Case Study 2: Multi-Signature Ownership Evolution}

\textbf{Innovation Problem:} A company evolved from single hardware wallet storage to multi-signature smart contract custody, fundamentally altering ownership verification requirements.

\textbf{Technical Implementation:} A single (hardware) wallet can sign a message with private keys to prove ownership, however, a smart contract does not have a private key to sign a message. Multi-signature Smart contracts implement complex governance requiring: (1) Source code analysis validating contract functionality; (2) Signatory verification ensuring keyholders possess claimed keys; (3) Governance mechanism assessment evaluating upgrade risks; (4) Emergency procedure evaluation understanding recovery mechanisms.

\textbf{Autoethnographic Analysis:} Implementing many self custody solutions across 12 different governance configurations over six years revealed the profound evolution of custody paradigms. When auditors requested "proof of ownership," I realized I couldn't simply demonstrate private key control—the very concept of ownership had evolved beyond traditional frameworks. Each multisig configuration presented unique verification challenges: 2-of-3 setups required different governance documentation than DAO (Decentralized autonomous organization) arrangements, and emergency recovery procedures varied dramatically based on keyholder distribution and organizational structure. The embodied experience of coordinating multiple keyholders for routine transactions—from simple transfers to complex smart contract interactions—revealed the inadequacy of existing audit procedures. This lived complexity spanning governance design trade-offs (security versus operational efficiency), technical custody evolution (key-based to smart contract-based), and control framework development could only be understood through direct implementation across multiple organizational contexts, not external observation~\cite{ellis2011autoethnography}.

\textbf{Auditing Challenges:} Auditors must develop technical expertise to analyze implementation and define ownership verification procedures. This requires understanding cryptographic primitives and network operations that traditional audits do not address~\cite{c4ccssa}.

\subsection{Case Study 3: Real-time Financial Reporting}

\begin{figure}[t]
    \centering
    \includegraphics[width=\columnwidth]{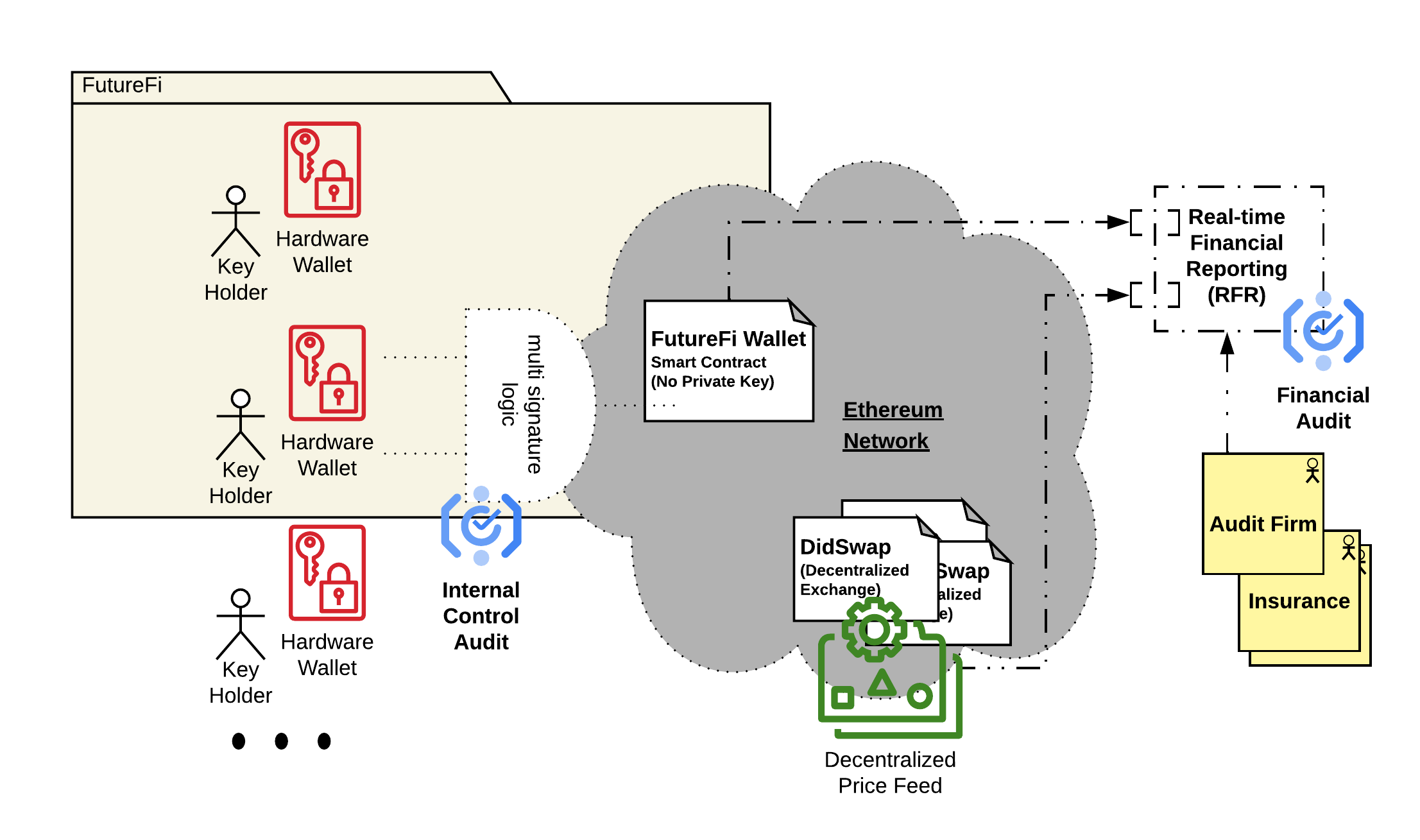}
    \caption{Case Study 3 - Real-time Financial Reporting (RFR): Blockchain transparency enables continuous audit verification but requires new control frameworks.}
    \label{fig:realtime_reporting}
\end{figure}

\textbf{Technical Innovation:} This experimental system enables auditors to verify asset quantities directly from blockchain nodes and obtain valuations from decentralized price feeds~\cite{eskandari2021oracles}, eliminating reliance on client-provided information (Figure~\ref{fig:realtime_reporting})~\cite{bakarich2020use}.

\textbf{Autoethnographic Analysis:} Proposing this solution during the last annual audit engagement covering \$100M+ in assets under management revealed that traditional auditors are not equipped with the technical expertise to feel comfortable with cryptographically verifiable data from the blockchain~\cite{vasarhelyi1991continuous}. The resistance I encountered when demonstrating real-time blockchain verification to our auditors was visceral and immediate. Despite providing mathematically verifiable proof of our holdings directly from blockchain nodes—data more reliable than traditional bank confirmations—they insisted on client-provided statements and manual reconciliation processes. This pattern repeated across multiple reporting periods: auditors would request CSV exports of on-chain data rather than verify it directly, creating additional points of failure and reducing audit quality. The cultural chasm between blockchain's inherent transparency and traditional audit culture's reliance on trusted intermediaries became most apparent during our year-end audits, where auditors questioned the integrity of immutable blockchain data while accepting mutable bank statements. The lived experience of managing financial reporting integration challenges—being simultaneously more transparent and less trusted—revealed the deep institutional inertia that autoethnographic analysis helps illuminate~\cite{ellis2011autoethnography}.

\textbf{Control Framework Requirements:} The approach requires new internal controls for: oracle security and redundancy, node infrastructure protection, automated reporting system integrity, and real-time monitoring mechanisms~\cite{wang2018designing}.

\section{Discussion and Implications}

\textbf{Methodological Contributions:} This research demonstrates that autoethnographic methodology is essential for understanding technical innovations that break existing paradigms. The lived experience of implementing blockchain systems provides insights that external analysis cannot generate~\cite{schmitz2019accounting,fisch2019initial}.

\begin{figure}[t]
    \centering
    \includegraphics[width=\columnwidth]{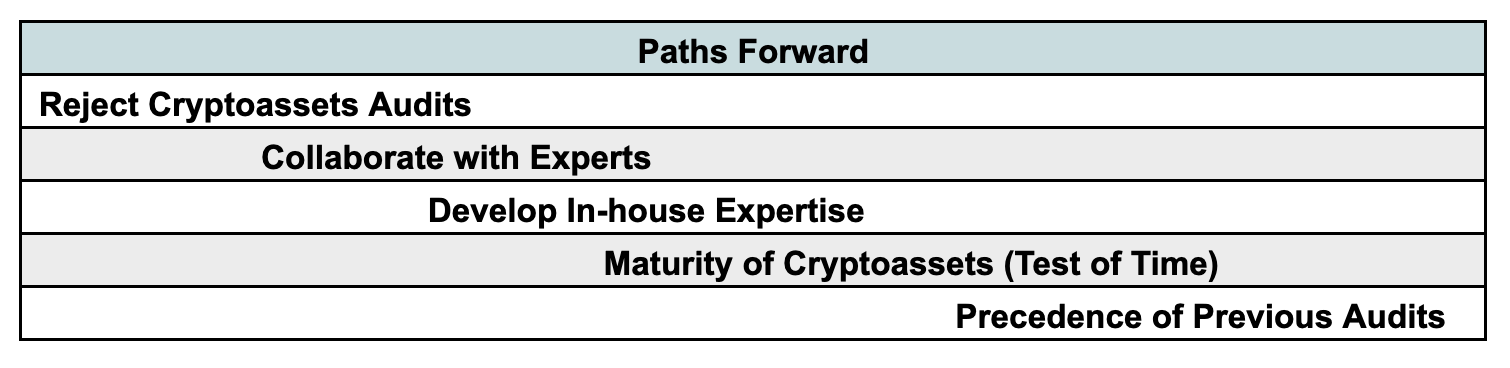}
    \caption{Paths Forward for Auditing Cryptoassets}
    \label{fig:paths_forward}
\end{figure}

Figure~\ref{fig:paths_forward} shows how auditing firms are evolving approaches to blockchain challenges. Our analysis reveals multiple pathways forward~\cite{han2023accounting}.

\textbf{Collaboration with Experts:} Successful implementations combine traditional audit expertise with blockchain security. Research shows relationship quality between auditors and IT specialists impacts audit evidence quality~\cite{bauer2019one,hirsch2020effect}.

\textbf{Developing In-house Expertise:} Major firms are creating specialized blockchain audit capabilities~\cite{regelbrugge2022internal,brody2022ey}. However, this requires significant investment and competition for limited blockchain expertise~\cite{cpab2024ai}.

\textbf{Test of Time:} As cryptoassets mature, more institutions recognize their relevance~\cite{bankofengland2022,kpmg2022rise,osfi2022interim}. Regulatory clarity is emerging, with frameworks developing for cryptoasset treatment~\cite{csa2022regulation,fras2023financial}.

\textbf{Standards Development:} Integration of traditional audit standards with cryptoasset security standards (like C4 CCSS) could create comprehensive frameworks~\cite{c4ccssa}. Proof-of-reserves methodologies are evolving into standardized procedures~\cite{kraken2022proof,dagher2015provisions}.

\section{Limitations and Future Work}

\textbf{Scope Limitations.} This analysis focuses on specific blockchain innovations experienced through professional implementation. Different technical pathways might reveal additional challenges requiring experimental approaches~\cite{coyne2017can}. The framework addresses traditional corporate entities holding cryptoassets but requires extension for decentralized autonomous organizations (DAOs).

\textbf{Methodological Extensions.} Future work should extend autoethnographic analysis to emerging innovations like zero-knowledge proofs, cross-chain protocols, and novel governance structures presenting new verification challenges~\cite{abreu2018blockchain}.

\textbf{Industry Validation.} Broader implementation across different auditing contexts will validate framework applicability and reveal additional technical dependencies requiring systematic approaches.

\section{Conclusion}

Blockchain innovations create fundamental challenges for traditional financial auditing requiring experimental approaches guided by technical understanding of the underlying technology and requirements. Token airdrops, multi-signature wallets, smart contracts, and real-time on-chain reporting represent paradigm shifts that existing frameworks cannot accommodate.

This research demonstrates the necessity of novel interdisciplinary methodologies for understanding technical innovations and breaking existing paradigms. Our approach provides a model for systematically analyzing emerging technologies requiring experimental verification, offering auditing firms pathways to address blockchain innovations rather than avoiding them entirely.

\section*{Acknowledgment}

This work was funded by Fundação para a Ciência e a Tecnologia (UID/00124/2025, UID/ PRR/124/2025, Nova School of Business and Economics DOI: https://doi.org/10.54499/ UID/00124/2025) and LISBOA2030 (DataLab2030 - LISBOA2030-FEDER-01314200). It also received support from the University Blockchain Research Initiative (UBRI), funded by the Ripple Impact Fund under grant 2022-310159 (251937), awarded to  the Data, Operations \& Technology Knowledge Center (DOT KC), Nova School of Business and Economics, Universidade NOVA de Lisboa, Campus de
Carcavelos, 2775-405 Carcavelos, Portugal.

\balance
\bibliographystyle{IEEEtran}
\bibliography{references/bibliography}

\end{document}